\documentclass[aps,prc,reprint,groupedaddress]{revtex4-1}

\usepackage{amsmath}

\usepackage{graphicx}
\usepackage{dcolumn}
\usepackage{bm}
\usepackage{cases}
\usepackage{longtable}

\begin{document}


\title{Shell evolution in neutron-rich nuclei: the single particle perspective}

\author{Weiqiang Ma}

\affiliation{Department of Applied Physics, Nanjing University of Science and Technology, Nanjing 210094, China}

\author{Yibin Qian}
\email{qyibin@njust.edu.cn}
\affiliation{Department of Applied Physics, Nanjing University of Science and Technology, Nanjing 210094, China}

\date{\today}

\begin{abstract}
	
The shell evolution has been studied extensively within the framework of interacting shell model, while the studies from the single particle viewpoint is relatively lacking or neglected. In particular, the isospin dependence of spin-orbit splitting has become increasingly important as $N/Z$ increases in neutron-rich nuclei. Following the initial independent-particle strategy towards explaining the occurrence of magic numbers, we have systematically investigated the isospin effect on the shell evolution of neutron-rich nuclei within the Woods-Saxon (WS) mean-field potential plus the spin-orbit term. It is found that new magic numbers $N = 14$ and $N =16$ may emerge in neutron-rich nuclei if one changes the sign of the isospin-dependent term in the spin-orbit coupling while the traditional magic number $N = 20$ may disappear. The magic number $N = 28$ is expected to be destroyed despite the sign choice of the isospin part in spin-orbit splitting, while $N = 50$ may disappear and $N = 82$ persists within the single particle scheme. Besides, an appreciable amount of energy gap appears at $N = 32$ and 34 in neutron-rich Ca isotopes. All these results are more consistent with those of the interacting shell model, when the sign of the isospin term of the WS potential is different from that of the corresponding spin-orbit coupling part. The present study may provide a more reasonable starting point for not only the interacting shell but also other nuclear many-body calculations towards the neutron-dripline.

\end{abstract}

\pacs{23.60.+e, 21.10.Gv, 21.10.Tg, 21.60.Gx}

\maketitle

\section{introduction}
Recently, owing to the development of both the experimental facilities and the computation capabilities, the study of shell evolution in neutron-rich nuclei has become an important and challenging subject. It provides a ground for understanding the composition of the nuclear force and how the elements in the universe are formed. The shell structure of the atomic nuclei is firstly explained by the independent particle model (IPM). The nuclear mean field potential plus the spin-orbit coupling correctly produces the so-called magic numbers such as 8, 20, 28, 50, 82, and 126 in stable nuclei\cite{a1, a2}. Inside the nucleus, the short-range nucleon-nucleon repulsion combined with the Pauli exclusion principle make the independent particle motion reasonable in the first approximation. However, as more nucleons are added into the valence orbits outside the closed shell, the residual two-body interaction between valence nucleons must be taken into account. The extended version of the IPM, namely the interacting shell model, is then proposed to tackle this complicated many-body problem \cite{a6,a7}. Meanwhile, the (effective) single-particle energy will change accordingly, leading that the magic number may emerge or disappear, namely the evolution of the shell closure. At present, it is believed that the monopole part of the nucleon-nucleon (NN) interaction (especially the tensor force effect) and the three-body interactions determine the shell evolution within the shell model context \cite{a8,a9,a10}. In this sense, new magic numbers such as $N=32$ and $N=34$ may emerge in calcium isotopes and $^{14,16,22,24}$O are expected to be four possible doubly-magic nuclei \cite{a10,a11}, which are also experimentally confirmed to some extent \cite{a21}.

Under the above mentioned framework, the shell evolution is interpreted and understood in view of the effective single particle energy (ESPE), which is composed of the naive single particle energy (SPE) and the shift generated by the residual interactions between valence nucleons. The latter one is actually being taken as the key physical quantity to detect the mechanics behind the shell evolution phenomenon. The former SPE comes from the calculated single-particle spectrum based on the mean filed potential felt by the individual nucleon. As compared to the monopole part of NN interaction or the three-body interactions, the attention placed on the SPE seems to be quite limited when considering the shell structural evolution. This is partly due to the fact that the contribution of SPE to the final ESPE is relatively small, and the SPE usually varies smoothly like for one isotopic chain \cite{a7}. Consequently, if one focuses on the change of the shell structure, the SPE can be only a basement and do not affect the variation of ESPE, namely the shell evolution. On the other hand, the conventional magic numbers regulated by the gaps of SPE spectrum have been very well known, while the deviation from this recognition is more appealing in terms of structural evolution. However, the isospin dependence has been widely absorbed into the single particle potential \cite{a14,Moller:2015fba}, resulting in the single particle spectrum as the starting point in all practical many-body methods. When it comes to the neutron-rich side of Segr\`{e} chart, the large neutron-proton asymmetry would enhance the isospin effect in the nuclear mean field. The SPE can be then not as previously expected, which may bring us a different mean field picture for the shell evolution of neutron rich nuclei.

The present study is aimed to clarify the role of the single particle potential played in the evolution of shell closure in particular for nuclei with extreme neutron-proton ratios. Recently, the shell structure of neutron rich nuclei has been somewhat analyzed by focusing on the isospin dependence of the spin-orbit splitting from the above phenomenological single-particle point of view \cite{a24}. Here we would like to make a systematical analysis on the shell evolution through tuning the isospin-related term of not only the mean-field WS potential but also the spin-orbit coupling. One can then answer the question: How does the weaken or enhancement of the spin-orbit splitting affect the shell evolution in neutron rich nuclei from a single particle perspective? We hope that the present SPE results can provide a more reasonable baseline for the probe into the shell evolution of exotic nuclei close to the neutron dripline. This paper is organized as follows. The construction of the effective WS single particle potential plus the isospin dependent term is introduced in Sec. II, and Sec. III presents the detailed discussions on the evolution of various shells for some typical or reported isotopes. A summary is given in the last section.

\section{Effective single particle hamiltonian and the choice of isospin dependent term}

Within the independent particle model, the key point is that the single nucleon motion is governed by the average field produced by all other nucleons. As is well known, the harmonic oscillator (HO) mean field plus the spin-orbit coupling term was the first successful mean field treatment, predicting the correct sequence of energy levels and the magic numbers \cite{a24}. However, nowadays the subjects of nuclear physics have been expanded far beyond the valley of stability into the broader region of nuclide chart, such as the neutron rich exotic nuclei. At that time, the continuum spectrum of the mean field potential appears to be indispensable, while the outer part of the HO potential cannot take this responsibility. Moreover, in reality, one may expect the spin-orbit (SO) coupling will be reduced for neutron dripline nuclei with a diffusive nuclear surface since the SO interaction is peaked at the surface of nuclei. This cannot be achieved by the gradient of the HO potential involved in the coefficient of the SO term either. Hence the more realistic Woods-Saxon potential, a common choice in modern nuclear theoretical techniques {a12,a13,a14}, is taken to describe the single particle shell structure.

After the subtle modifications via considering the reduced mass and the isospin symmetry \cite{a14}, the total effective single-nucleon hamiltonian reads as
\begin{equation}\label{eq1}
H=\frac{p^2}{2\mu}+V(r)+V_{c}(r)+\frac{\hbar^2}{2\mu c^2r}\left(\frac{\partial}{\partial r}\widetilde V(r)\right) \textbf{L}\cdot\textbf{S}
\end{equation}
The first term is single-nucleon kinetic energy, and $\mu$ is the reduced mass of a nucleon-core system. Based on the lowest order isospin invariant, the effective nuclear potential $V(r)$, related with the scalar product of the isospin of the nucleon $\textbf{t}$ and the core $\textbf{T}^{'}$, is taken as \cite{Lane}
\begin{align}\label{eq2}
	V(r)=-Vf(r,R,a)\quad,\quad\quad\quad
	V=V_0\left(1-\frac{4\kappa}{A}\left\langle\textbf{t}\cdot\textbf{T}^{'}\right\rangle\right),
\end{align}
where $V_0$ is the strength parameter of nuclear potential. The coefficient $\kappa$ regulates the isospin dependent term of the nuclear potential, tuning the depth of the nuclear potential as well. By combining the relationship $\textbf{t}+\textbf{T}^{'}=\textbf{T}$ and the assumption that the isospin number is $T=|T_z|=|N-Z|/2$ for the ground state of one nucleus, the behavior of $-4\left\langle\textbf{t}\cdot\textbf{T}^{'}\right\rangle$ is then determined as
\begin{center}
\begin{equation}\label{eq6}
 -4\left \langle\textbf{t}\cdot\textbf{T}^{'}\right\rangle =
    \begin{cases}
       3\ & N = Z \\
       \pm(N-Z+1)+2\ & N \geq Z \\
       \pm(N-Z-1)+2\ & N \leq Z.
    \end{cases}
\end{equation}
\end{center}
Here the upper and lower signs denote the proton and neutron, respectively. Such a modified isospin dependence, introduced in the nuclear mean-field potential, can lead to interestingly different spectrums in light nuclei around $N=Z$ \cite{a14}. The coulomb potential $V_c(r)$, corresponding to a nucleon electromagnetically interacting with a uniformly charged sphere of radius $R_c$, is given by
\begin{equation}\label{eq3}
V_c(r) = \frac{Z^{'}e^2}{4\pi\varepsilon_0}
\begin{cases}
\dfrac{3-\left(\frac{r}{R_c}\right)^2}{2R_c}\ & r \leq R_c \\
\frac{1}{r} & r \geq R_c,
\end{cases}
\end{equation}
where $Z^{'}$ is the proton number of the core nucleus. As for the last spin-orbit coupling term, the potential $\widetilde V(r)$ is supposed to be not equal to the original potential $V(r)$ along with a different form factor \cite{a14}. Consequently, the form of $\widetilde V(r)$ is constructed as
\begin{align}\label{eq4}
	\widetilde{V}(r)=\widetilde{V}f(r,R_{so},a_{so})\quad,\quad\quad\quad
	\widetilde{V}=\lambda V_0\left(1-\frac{4\kappa_{so}}{A}\left\langle\textbf{t}\cdot\textbf{T}^{'}\right\rangle\right),
\end{align}
The characteristic function $f(r,R,a)$, as suggested by Woods and Saxon \cite{a12}, is in the Fermi form,
\begin{equation}\label{eq5}
f(r,R,a)=\left[1+\exp\left(\frac{r-R}{a}\right)\right]^{-1},
\end{equation}
where the size parameter $R$ and surface diffuseness $a$ are to be determined. Following the conventional choice of WS potential parameterization, the size and diffuseness parameters are regulated by $R = R_c = R_0A^{\frac{1}{3}}$, $R_{so} = R_{0,so}A^{\frac{1}{3}}$, and $a=a_{so}$. As mentioned before, the isospin dependence in the mean-filed nucleon-nucleon potential appears to be increasingly important towards unstable nuclei with extreme neutron-proton ratio. In this sense, to probe into the evolution of single particle shell, one should pay special attention to the key isospin-related parameters $\kappa$ and $\kappa_{so}$ in the strength of central and spin-orbit interactions respectively. In order to systematically investigate the isospin effect in the SPE evolution, three kinds of calculations with $\kappa_{so}=\pm\kappa, 0$ are performed here with the GAMOW code \cite{a16}. Similar to the global optimization procedure in Ref.~\cite{Qi}, all the above parameters in the single particle hamiltonian are adjusted to single particle and single hole states in the vicinity of the doubly magic nuclei
$^{16}_{8}$O, $^{40}_{20}$Ca, $^{48}_{20}$Ca, $^{100}_{50}$Sn, $^{132}_{50}$Sn, and $^{208}_{82}$Pb, as listed in Refs.~\cite{a14,isakov2002}. In Table~\ref{tab:table1}, the parameter set obtained in this study are presented as compared with other theoretical results for different cases of $\kappa$ and $\kappa_{so}$.

\begin{widetext}

\begin{table}
	\caption{\label{tab:table1}Woods-Saxon potential parameters obtained by fitting to the available single-particle and single-hole states around doubly-magic nuclei with the restriction $\kappa_{so}=\pm\kappa,0$ and comparison with existing parameterizations.}
	\begin{ruledtabular}
		\begin{tabular}{cccccccc}
			\quad&$V_0(Mev)$&$r_o(fm)$&$r_{0,so}(fm)$&$a=a_{so}(fm)$&$\lambda$&$\kappa$\\
			\hline
			\quad&51.47&1.278&1.165&0.654&23.165&0.644&$\kappa_{so}=\kappa$\\
            \cite{blom1960}&51&1.27&1.27&0.67&32.13&0.647&$\kappa_{so}=\kappa$\\
            \cite{Dudek}&49.6&1.374(n)/1.275(p)&1.31(n)/1.32(p)&0.7&35(n)/36(p)&0.86&$\kappa_{so}=\kappa$\\
			\quad&51.25&1.283&1.076&0.637&20.716&0.640&$\kappa_{so}=-\kappa$\\
            \cite{a24}&50.92&1.285&1.146&0.691&24.07&0.644&$\kappa_{so}=-\kappa$\\
			\quad&51.40&1.279&1.129&0.647&22.112&0.643&$\kappa_{so}=0$\\
			\cite{a14}&52.06&1.26&1.16&0.662&24.1&0.639&$\kappa_{so}=0$\\
		\end{tabular}
	\end{ruledtabular}
\end{table}

\end{widetext}
\section{NUMERICAL RESULTS AND DISCUSSION}
The shell evolution in exotic nuclei has received special attention for decades \cite{a10,a17,a18,a19,a20,Momiyama}, such as the emergence of the new magic numbers $N = 14$, 16, 32, and 34 and the disappearance of traditional magic number $N = 20$ \cite{a17,a18,a19,a20}. Such exotic phenomena are supposed to be attributed to the monopole part of the nucleon-nucleon interaction, especially the tensor force \cite{a22,Zalewski}, within the interacting shell model. In the present study, we attempt to systematically analyze these expected or reported evolutions of shell structure from the identical particle viewpoint. This procedure can not only directly check the isospin effect on the SPE evolution, but also provide a more comprehensive benchmark for further microscopic studies on shell evolution in view of the ESPE variation. The emergence of new magic numbers and the evolution of the traditional shell closure will be specifically studied in the following subsections.

\subsection{New magic numbers $N =$ 14 and 16 in Oxygen isotopic chain}

The $N =$ 14 shell gap is formed between the $1d_{5/2}$ and $2s_{1/2}$ orbits, while the $N =$16 shell gap is located between the $2s_{1/2}$ and $1d_{3/2}$ orbits. The evolution of ESPE of the $1d_{5/2}$, $2s_{1/2}$, and $1d_{3/2}$ neutron orbits caused by the monopole interactions between valence nucleons indicated $^{22,24}$O are quasi-doubly-magic nuclei in Oxygen isotopic chain\cite{a11}. To be specific, the $N = 14$ gap is created by the filling of 6 neutrons in the $1d_{5/2}$ orbit. As the monopole interactions between neutrons in $1d_{5/2}$ orbit are globally attractive, the single-particle energy of the $1d_{5/2}$ orbit gains extra binding energy. Meanwhile the single-particle energy of the $2s_{1/2}$ orbit moves upwards, due to the monopole interactions between neutrons in $1d_{5/2}$ and $2s_{1/2}$ orbits are slightly repulsive. As the neutrons start to fill the $2s_{1/2}$ orbit, a large shell gap $N = 16$ emerges due to the presence of monopole interactions between neutrons in $2s_{1/2}$ orbit. In this paper, the single-particle energies of the $1d_{5/2}$, $2s_{1/2}$, and $1d_{3/2}$ neutron orbits are calculated from single-particle viewpoint for the oxygen isotopic chain. To get a qualitative idea on the role played by isospin term of single-particle SO potential in shell evolution of neutron-rich nuclei, the calculations with $\kappa_{so} = \pm\kappa$ and $\kappa_{so} = 0$ are preformed, respectively. In each calculation, the corresponding Woods-Saxon parameter set in table 1 is used.

Looking at the the calculated single-particle energy in Fig. 1, one can notice that if one takes $\kappa_{so}=+\kappa$ (red dotted line) weakening the strength of spin-orbit potential in neutron-rich nuclei, the single-particle energies of $1d_{5/2}$ and $2s_{1/2}$ orbits stay roughly constant as the neutron number increases. The gap between them is about 1.7 Mev. Meanwhile, the single-particle energy of $1d_{3/2}$ decreases rapidly. In fact, as more and more neutrons fill into the $1d_{5/2}$ orbit, this orbit should lose its energy rapidly due to the attractive monopole interaction, while the $2s_{1/2}$ orbit moves upward caused by the slightly repulsive monopole interaction. Such a case can be better explained, if one takes $\kappa_{so}=-\kappa$ (black dotted line), the gap between the $1d_{5/2}$ and $2s_{1/2}$ orbits then increases significantly in neutron-rich nuclei. It goes from 1.6 Mev in $^{16}$O to 2.9 Mev in $^{30}$O, as companied by an obvious slope of single-particle energy curve of $1d_{5/2}$ orbit. This slope is almost a constant as shown in the figure, while it is expected to be smaller as the neutrons start to fill $2s_{1/2}$ orbit (above $N=14$) according to the result of the interacting shell model \cite{a10}. Besides, the $2s_{1/2}$ orbit is independent on the spin-orbit splitting from the single particle model perspective. Meanwhile, the $1d_{3/2}$ orbit gradually loses its energy slowly in neutron-rich nuclei, implying a less bound state. These present results, with $\kappa_{so}=-\kappa$, are more consistent with the results of the interacting shell model. The reason may be that the effect of the monopole interactions are accidentally partly replaced by the enhancement of spin-orbit splitting. This point will be further discussed in the following cases. By the way, the situation with $\kappa_{so}=0$ (blue solid line), as expected, is somewhere in the middle of those two extreme cases $\kappa_{so}\pm \kappa$. Of course, we have to say that the results obtained in this paper only show that $N = 14$ may be a new magic number because the gap in neutron-rich nuclei is not large enough. Besides, the gap between the $1d_{3/2}$ and $2s_{1/2}$ orbits decreases with the increasing of the neutron number $N$ no matter which form of $\kappa_{so}$ is concerned. The present results, therefore, do not support the emergence of the $N = 16$ magic number to a large extent in neutron-rich nuclei. However, if one takes $\kappa_{so}=-\kappa$, the support could be more positive. The above discussions are based on the systematics of one isotopic chains, the conjecture can be more complete by comparing with the results of isotones as shown in the next part.

\begin{figure}[h]
	\includegraphics[width=2.8in,height=2.8in]{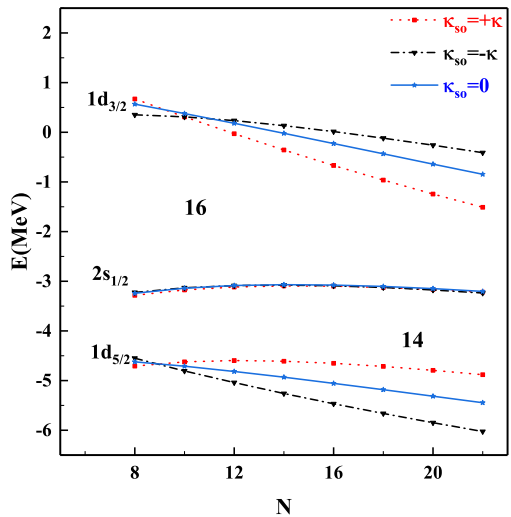}
	\caption{\label{fig:epsart} Evolution of the neutron single-particle energies of $1d5/2$, $2s1/2$, and $1d3/2$ orbis in O isotopes as a function of the neturon number $N$ for calculations with $\kappa_{so}=\pm\kappa$ and 0.}
\end{figure}

\subsection{Comparative analysis of magic numbers $N=16,20$ in the isotones}

As compared with the monopole part of neutron-neutron interaction mentioned in subsection A, the monopole part of neutron-proton interaction, especially the tensor force, plays a major role in the appearance of $N = 16$ magic number and the disappearance of $N = 20$ shell closure \cite{a10, a19, a22, a23}. The major $N = 20$ shell comes from the large gap between the $1d_{3/2}$ and $1f_{7/2}$ orbits, and the gap between the $2s_{1/2}$ and $1d_{3/2}$ orbits is supposed to produce one possible closed shell $N=16$. The magic number $N=$20 can arise naturally from the single particle viewpoint, while the new shell structure can be formed due to a sizable spin-isospin coupling (tensor force) in exotic nuclei. In the framework of the interacting shell model, as the protons occupy the $1d_{5/2}$ orbit, the monopole interactions between protons of the $1d_{5/2}$ orbit and neutrons of the $1d_{3/2}$ orbit can lower the neutron $1d_{3/2}$ orbit with respect to the $2s_{1/2}$ orbit. In other words, the $1d_{3/2}$ neutron orbit moves upward relative to the $2s_{1/2}$ orbit when protons are taken out from $1d_{5/2}$ orbit. Therefore, the $N = 16$ shell gap increases, while $N = 20$ shell gap decreases due to the fact that the effect of interactions between protons in $1d_{5/2}$ orbit and neutron in $1_{7/2}$ orbit is relatively weaker.
\begin{figure}[h]
	\centering
	\includegraphics[width=2.6in,height=2.6in]{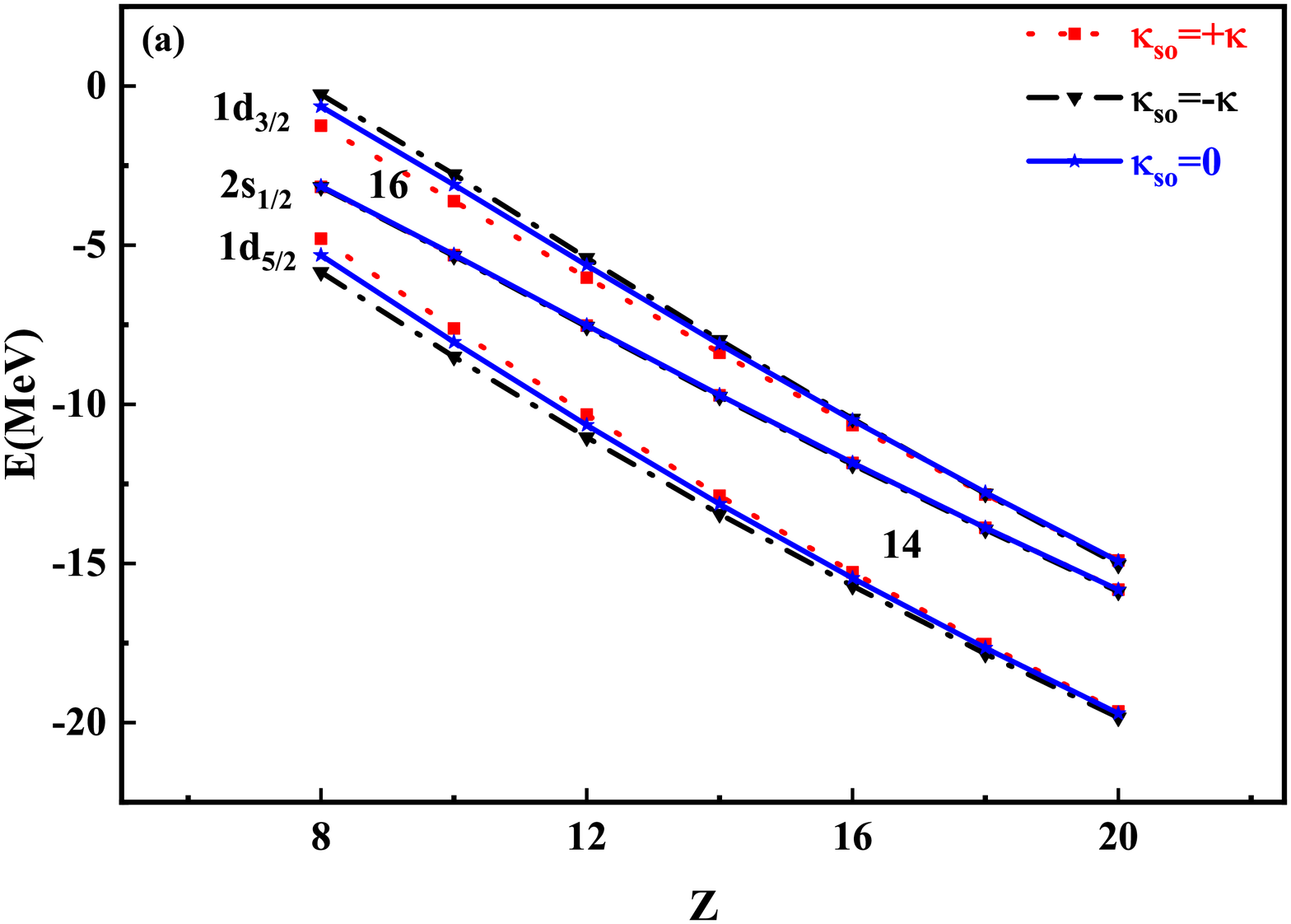}
	\includegraphics[width=2.6in,height=2.6in]{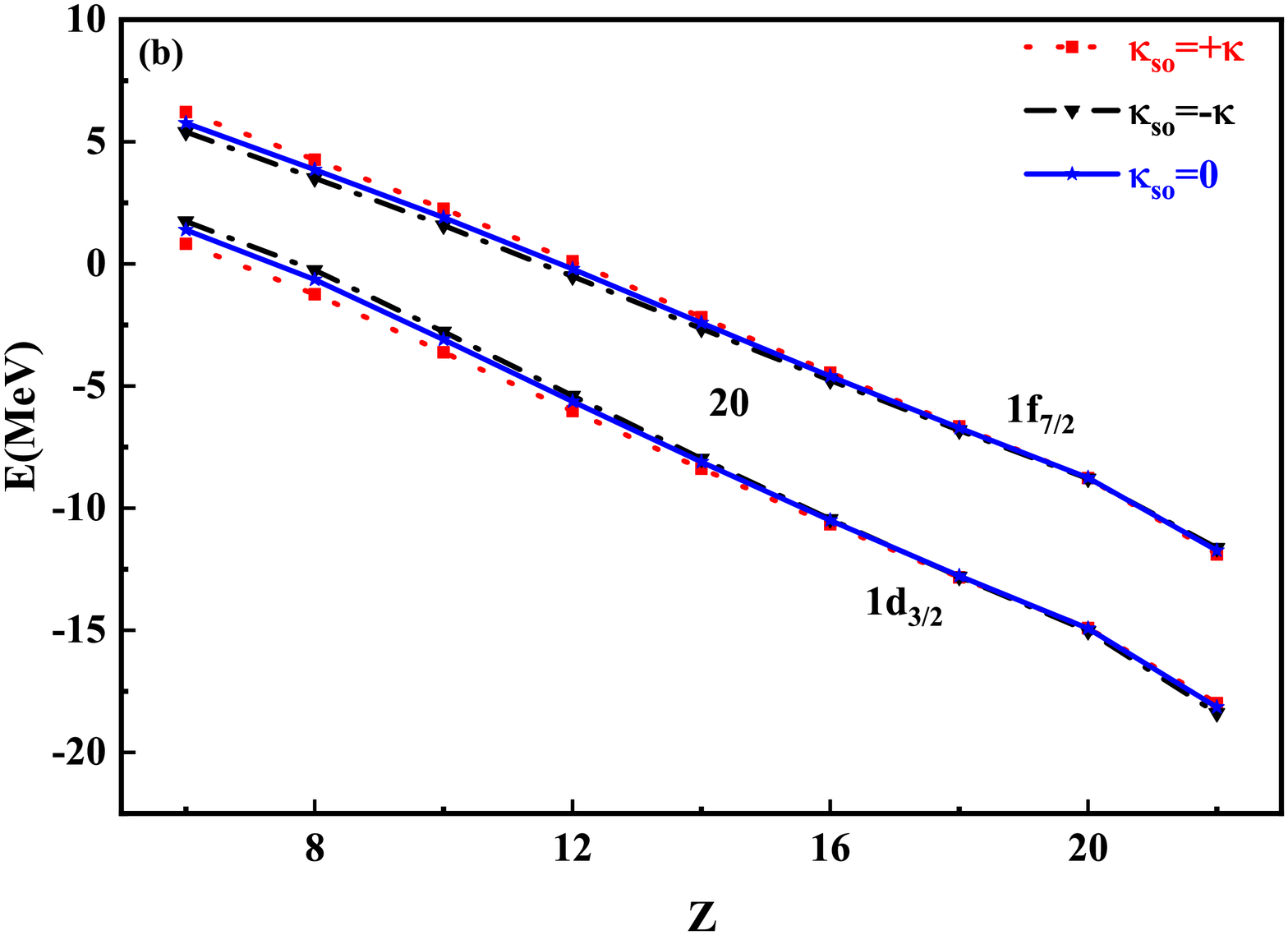}
	\caption{\label{fig2} SPEs of $1d_{3/2}$, $2s_{1/2}$ and $1d_{5/2}$ orbits in the panel (a) and of $1f_{7/2}$ and $1d_{3/2}$ in the panel (b) versus the proton number $Z$ with fixed 20 neutrons.}
\end{figure}

At present, the neutron number of target nuclei is fixed at $N=20$, and we have then performed a systematical calculation on the SPEs of the involved orbits in the evolution of the $N=16,\ 20$ shell with the filling of the valence protons. Firstly, we concentrate on the evolution of $N = 20$ shell gap as shown in Fig.~\ref{fig2} (b). The gap between $1d_{3/2}$ and $1f_{7/2}$ orbits remains a constant in the whole picture if one takes $\kappa_{so}=+\kappa$ (red dotted line). In fact, however, it should decrease as the proton number decreases when the number of protons is less than 16 due to a stronger force caused by the protons of $1d_{5/2}$ orbit acting on neutrons of $1d_{3/2}$ orbit than on those of $1f_{7/2}$ orbits according to the interacting shell-model analysis \cite{a10}. This point could be different if one takes $\kappa_{so}=-\kappa$ (black dotted line). In detail, by looking at the variation of the calculated single particle energy in Fig.~\ref{fig2} (b), one can see that the $N = 20$ gap indeed decreases when the proton number is less than 16 despite its slow speed. For illustration, the $N=20$ gap is 4.2 Mev for $^{28}$O (corresponding to the beginning of Fig.~\ref{fig2}), and appears to be quite sizable as compared with other theoretical results \cite{a14,a24}. Next, let us place our attention to the evolution of $N = 16$ gap as shown in Fig.~\ref{fig2} (a). The gap slowly increases as the proton number decreases even though one takes $\kappa_{so}=+\kappa$ (red dotted line). It goes from 0.9 MeV in $^{40}$Ca to 2.0 MeV in $^{28}$O. This can gap increase to 3.0 MeV in $^{28}$O if one takes $\kappa_{so}=-\kappa$ (black dotted line). Particularly, the slope of single-particle energy curve of $1d_{3/2}$ is greater when $Z<16$. This is consistent with the conclusion that there are strong attractive monopole interactions between the nucleons of $l+1/2$ and $l-1/2$ orbits \cite{a10, a19}. As the protons are taken out of the $1d_{5/2}$ orbit, the $1d_{3/2}$ neutron orbit moves upward. This indicates that the new magic $N = 16$ may appear, while the disappearance of magic number $N = 20$ is not obvious.

\subsection{Appearance of N = 32 and 34 magic number in the Ca isotonic chain}

\begin{figure}[h]
	\centering
	\includegraphics[width=2.6in,height=2.6in]{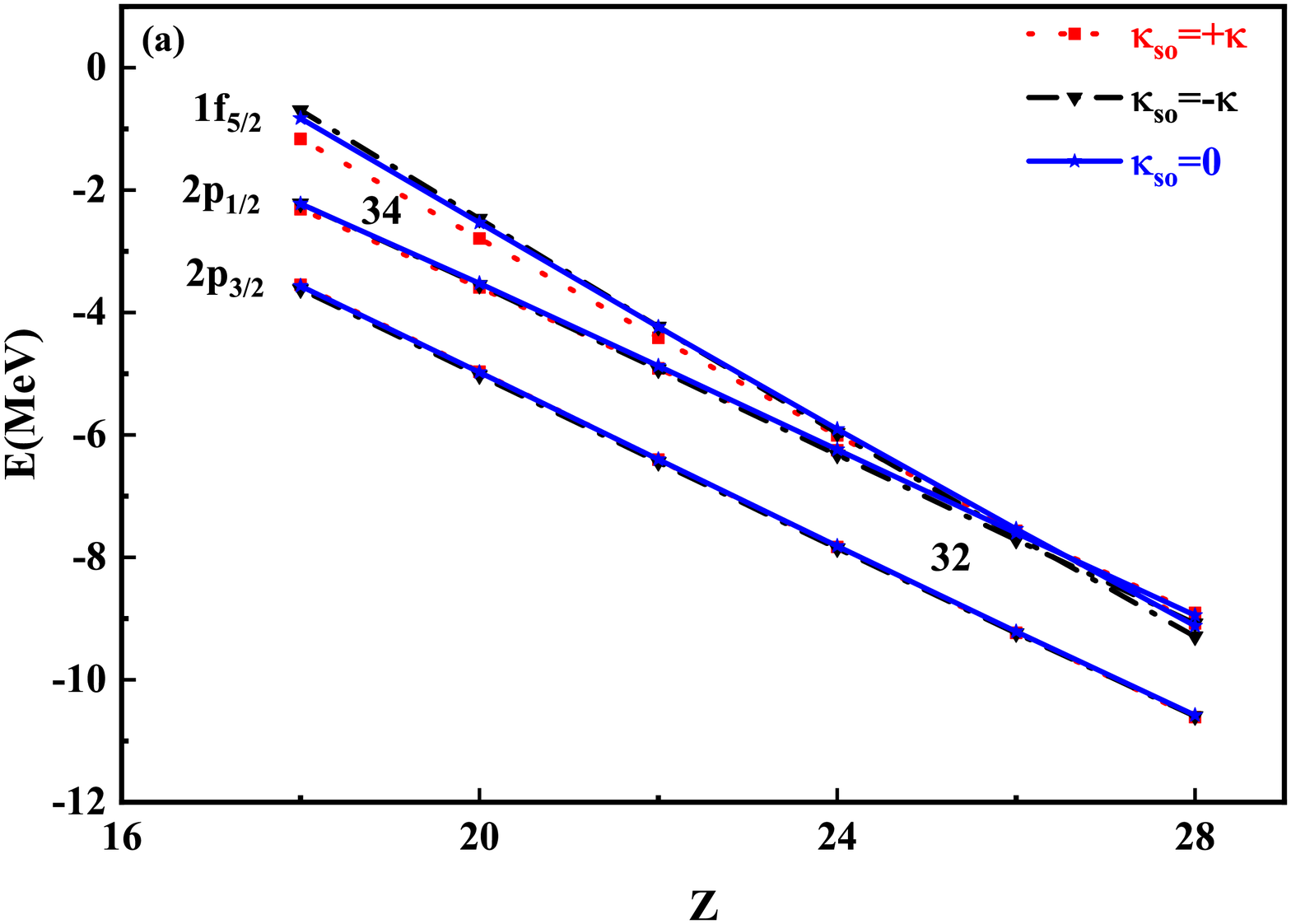}
	\includegraphics[width=2.6in,height=2.6in]{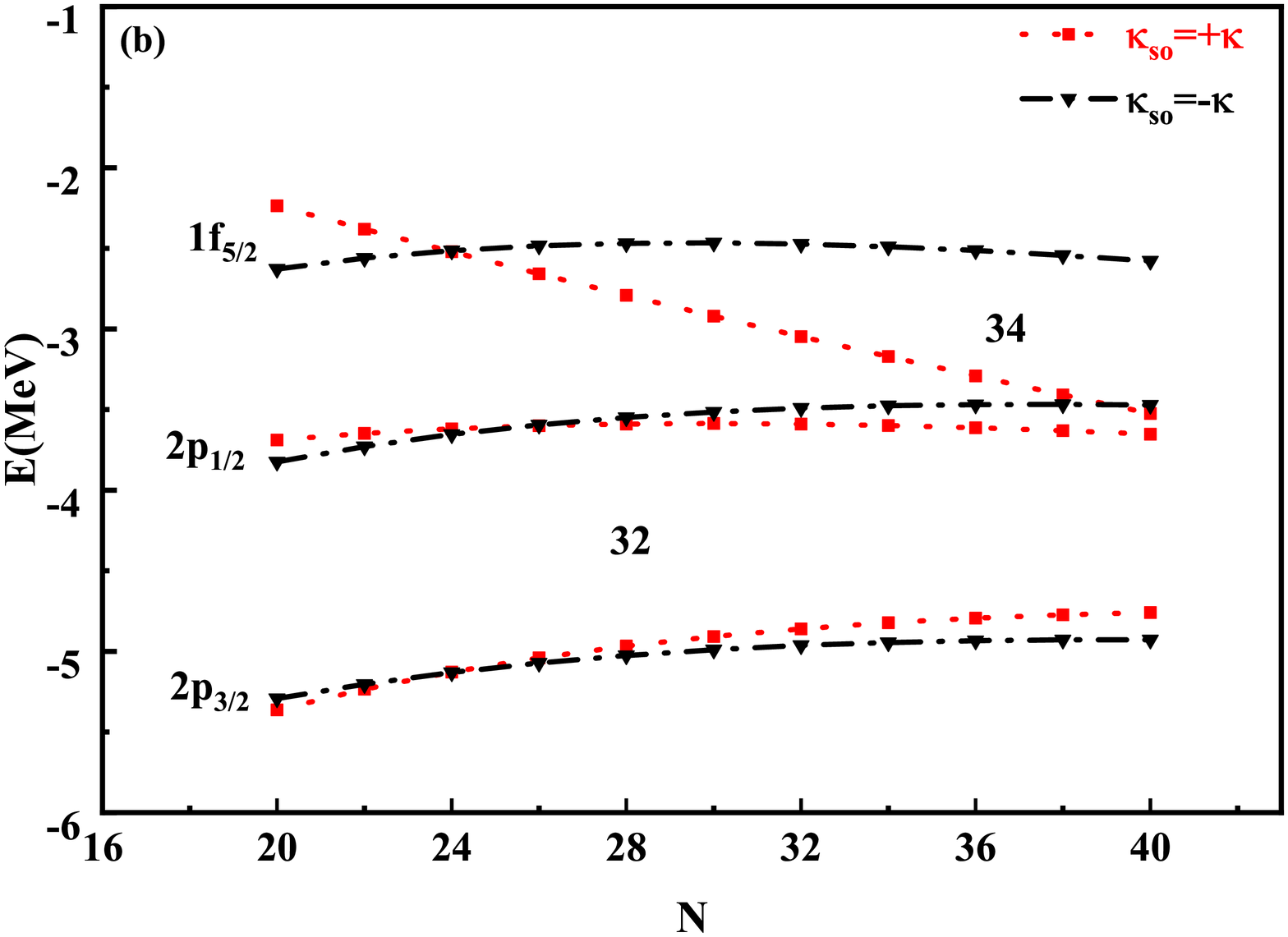}
    \caption{\label{fig3} Calculated single-particle energies of the $2p3/2$, $2p1/2$, and $1f5/2$ neutron orbits as function of (a) the proton ($N=28$ isotones) and (b) neutron (calcium isotopes) number with the different choices of isospin-related parameters $\kappa_{so}=\bm\kappa$ and 0.}
\end{figure}

Another important report of shell evolution is that $N = 32$ and 34 could be new magic numbers, based on the unexpected behaviors of the excitation energies of first 2$^+$ states and large charge radii in the Ca isotopes and their neighbors \cite{a21,Steppenbeck,Ruiz2016}. The $N = 32$ and $N=34$ shell gaps are formed between the $2p_{3/2}$ and $2p_{1/2}$ orbits and between the $2p_{1/2}$ and $1f_{5/2}$ orbits, respectively. The strong attractive interaction between the proton in the $1f_{7/2}$ orbit and the neutron in the $1f_{5/2}$ orbit is expected to play a crucial role in this case, which is analogous to the mechanism leading to the production of the $N = 16$ new magic number \cite{a10}. As the eight protons in $^{56}$Ni are taken away from the $1f_{7/2}$ orbit, new magic numbers $N = 32$ and 34 appear in calcium isotopes. In this paper, these two new magic numbers are studied by calculating the SPE of related orbits from the identical particle viewpoint. At the left panel of Fig.~\ref{fig3}, the varying SPE patterns of the $2p_{3/2}$, $2p_{1/2}$, and $1f_{5/2}$ orbits are presented as the proton number decreasing from 28 to 20 with the fixed $N=28$. One can find that all orbits, in Fig.~\ref{fig3} (a), move upward rapidly with the decreasing of the proton number and the $N = 34$ shell gap increases despite the choices of the $\kappa$ or $\kappa_{so}$. Additionally, this gap increases a little faster if one takes $\kappa_{so}=-\kappa$ (black dotted line). As a typical point, the $N = 34$ shell gap is about 1.6 MeV in $^{48}$Ca when $\kappa_{so}=-\kappa$. Meanwhile, the gap between the $2p_{3/2}$ and $2p_{1/2}$ orbits roughly keeps a constant (about 1.4 MeV) in all cases. These are consistent with the judgement on the shift of the $1f_{5/2}$, $2p_{1/2}$, and $2p_{3/2}$ orbits due to the tensor- and central- force monopole contributions \cite{a10}. Particularly, the $1f_{5/2}$ orbit gradually elevates above the $2p_{1/2}$ orbit leading to the final emergence of the new magic number $N = 32$, which takes place when the two protons are moved away from the $1f_{7/2}$ orbit, as shown in the part (a) of Fig.~\ref{fig3}.

Through the above systematical discussion on isotones with $N$, one may conclude that the effect of the monopole part of neutron-proton interaction on evolutions of the $N=32$ and 34 shell are manifested to some extent by enhancing the strength of the SO splitting via the choice of $\kappa_{so} = -\kappa$. Furthermore, one may expect to see what will happen when the $2p_{1/2}$, $2p_{3/2}$, and $1f_{5/2}$ orbits start to be filled by the valence neutrons. The interacting shell model studies show that the monopole part of the neutron-neutron interaction can be repulsive, and its effect on the $N=32$ and 34 shell evolution is limited by comparing the SPEs of involved orbits in $^{48}$Ca and those in $^{54}$Ca. Here a similar conclusion can be obtained from the single particle viewpoint, as shown in the part (b) of Fig.~\ref{fig3}. By choosing the $\kappa_{so}=\kappa$ and $\kappa_{so}=-\kappa$ respectively, the calculated SPEs of $2p_{3/2}$, $2p_{1/2}$, and $1f_{5/2}$ orbits are presented versus the neutron number of calcium isotopes, where one can find the evident distinction at the $N=34$ subshell gap for the two choices of $\kappa$ and $\kappa_{so}$. The near degeneracy between the $1f_{5/2}$ and $2p_{1/2}$ orbits in neutron-rich is eliminated if one takes $\kappa_{so}=-\kappa$ for the SO potential. Moreover, under such a mean-filed potential, the energy gap between those single orbits change quite gently with the increasing of the neutron number, which is consistent with the aforementioned comment about the limited influence of neutron-neutron monopole interaction on the formation of the $N=32$ and 34 subshell gaps. For instance, the $N=32$ gaps of $^{48,54}$Ca are both 1.5 MeV. Based on the above analysis, new magic numbers $N = 32$ and 34 can be emerged by strengthening of the single particle SO potential via the choice of $\kappa_{so}=-\kappa$.

\subsection{Traditional magic number N = 28, 50, and 82}

\begin{figure}
	\centering
	\includegraphics[width=2.0in,height=2.0in]{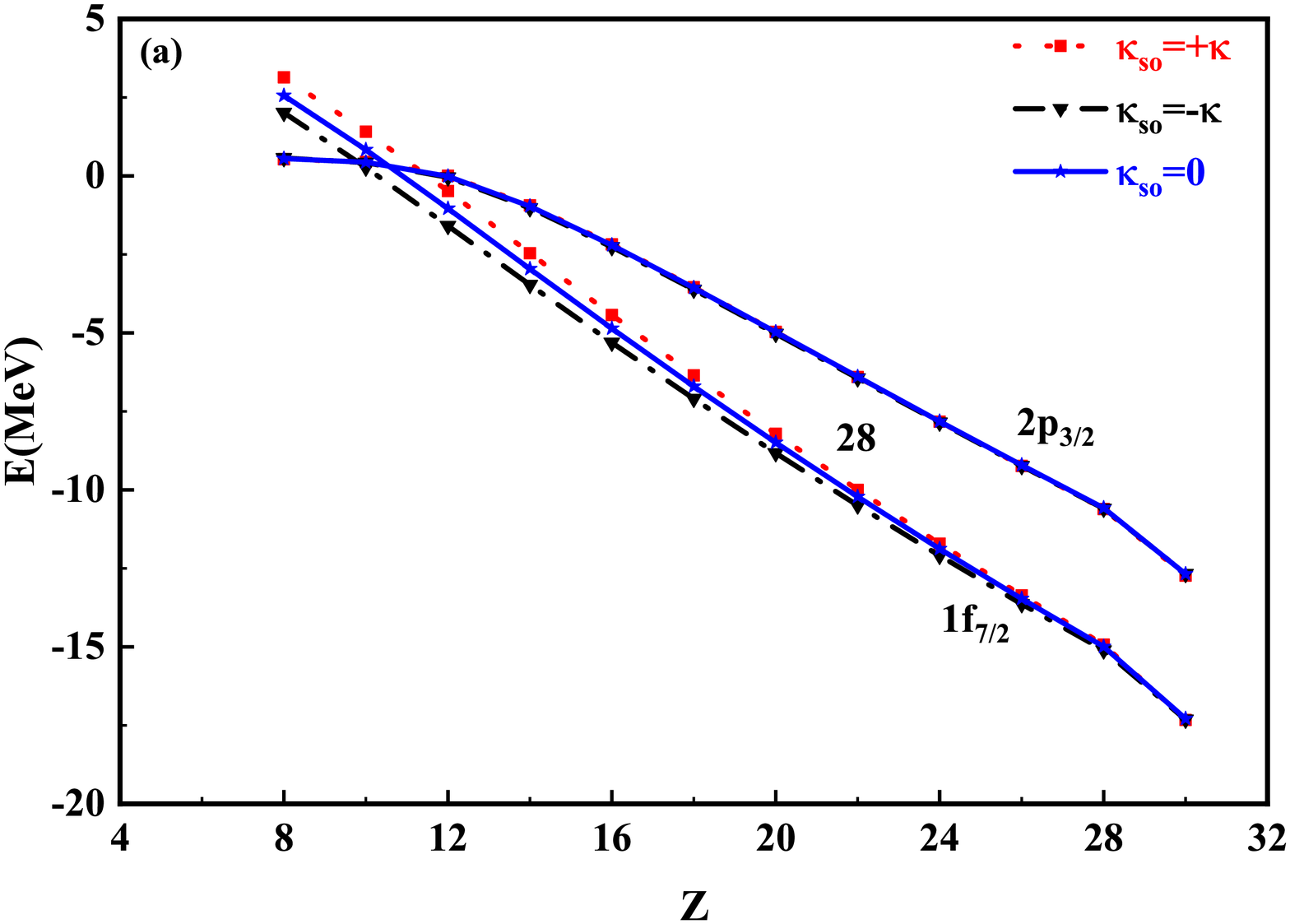}
	\includegraphics[width=2.0in,height=2.0in]{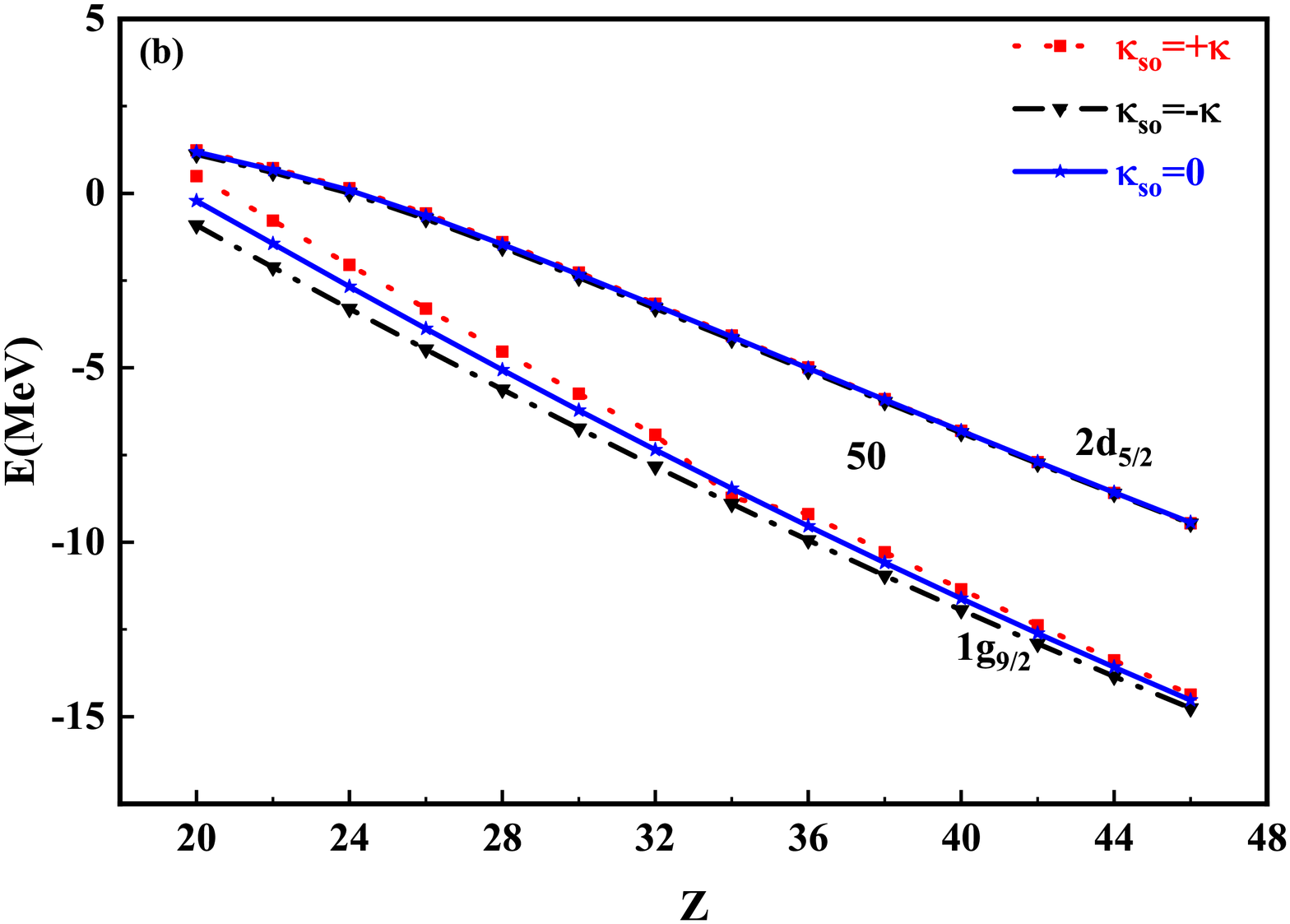}
	\includegraphics[width=2.0in,height=2.0in]{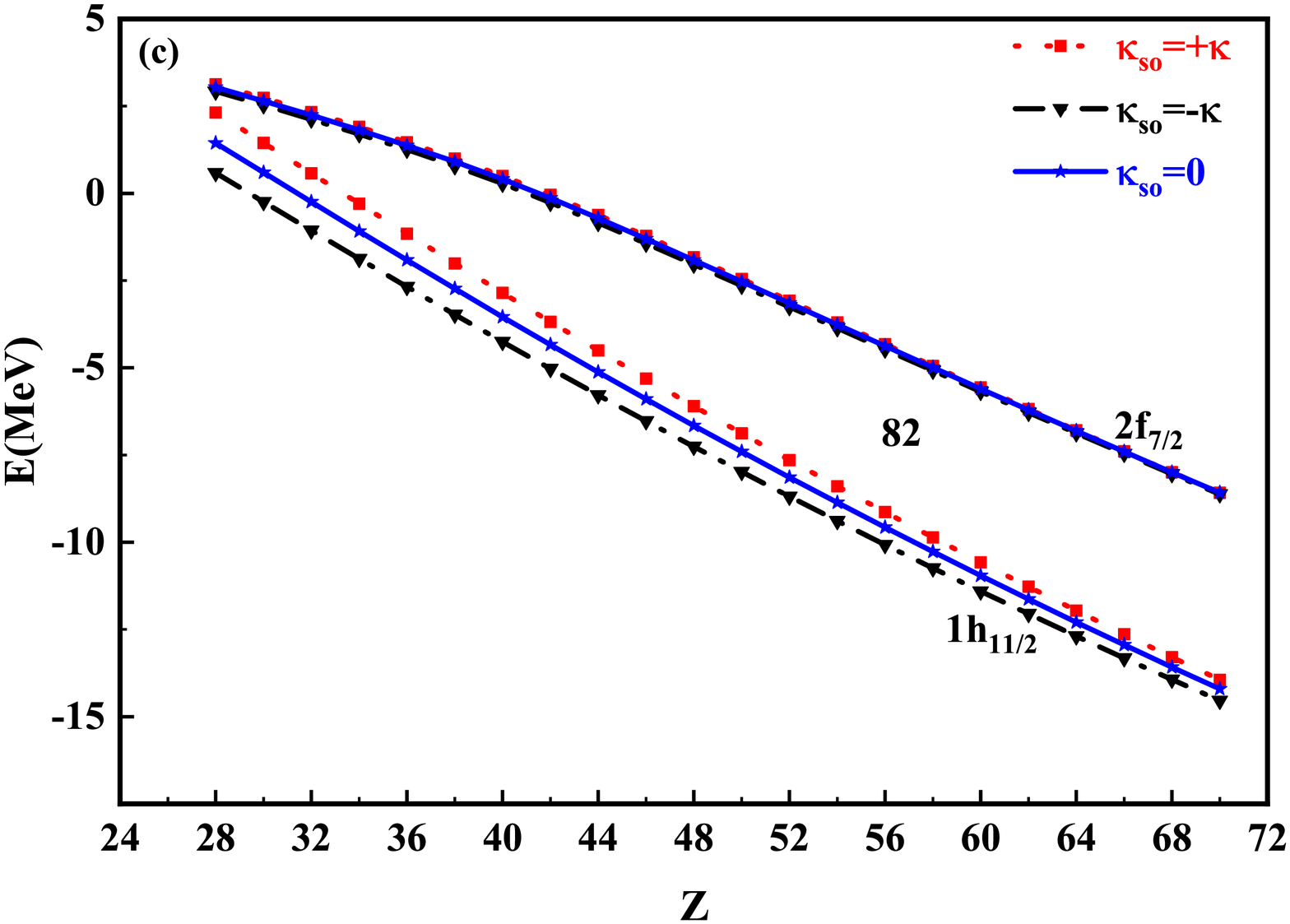}
	\caption{ \label{fig4} Same as previous figures but for the evolution of the $N=28$, 50, and 82 gaps as function of the proton number $Z$}
\end{figure}

Last but not least, we have also studied the evolutions of the traditional neutron magic numbers $N = 28$, 50, and 82 in neutron-rich nuclei. The investigation on these shell evolutions is significant for realizing the fundamental properties of nuclei, such as the nuclear force and the nuclear shape coexistence \cite{a25,a26,a27}. However, the problem would be much more complicated due to a large number of involved nucleons, leading that more factors need to be taken into account like the nuclear deformation \cite{a28,a29,a30,a31}. Although some experimental and theoretical results have been obtained, more details are still lacking. In the present study, we attempt to provide valuable information on this topic from the single particle perspective. Fig.~\ref{fig4} presents the varying SPEs of related orbits with the proton number for the $N=28$, $N=50$ and $N=82$ isotones, respectively. It is found that the conventional magic number $N=28$ is destroyed towards the neutron-rich region (see the left part (a) of Fig.~\ref{fig4}) regardless of the choice of the single particle SO splitting. As compared, when it comes to the neutron-dripline, there will be still energy gap between the $1g_{9/2}$ and $1d_{5/2}$ orbits as shown in Fig.~\ref{fig4} (b), corresponding to the $N=50$ shell closure. However, the gap would be quite small in contrast with that in the neutron-deficient side of nuclide chart. For example, the energy gaps between the $1g_{9/2}$ and $1d_{5/2}$ orbits, of the extreme neutron-rich nucleus $^{70}$Ca, are separately 0.7 MeV and 2.1 MeV for the cases $\kappa_{so}=\kappa$ and $\kappa_{so}=-\kappa$. The similar situation occurs at the $N=82$ shell gap, while the gap could be slightly larger for the extreme neutron-rich nuclei, implying the possible persistence of the magic number $N=82$. A noteworthy fact is that in the three parts of Fig.~\ref{fig4}, when all protons in the proton shell closure are taken out, the energy gap between neutron orbits begins to decrease more drastically. As seen from the figure, the $N = 28$ shell gap decreases faster when $Z<20$. Besides, the $N = 50$ and 82 shell gaps also drop quickly when $Z<28$ and $Z<50$, respectively. These may indicate that the interaction between protons and neutrons across the major shell closure may have an important impact on the shell evolution phenomenon.

%
\section{Conclusion}

Within the Woods-Saxon mean-field potential plus the spin-orbit coupling, we have systematically studied the robustness of the traditional magic numbers and the possible emergence of new shell closure in neutron-rich nuclei from a single particle perspective. By refining the isospin related term in the SO coupling potential, it is found that the well-known magic number $N=28$ will disappear towards the extreme neutron-proton ratio, such as the light nuclei with $Z<12$. As compared, the $N=20$ neutron shell closure tends to be eroded in a very small pace or may say survive when it comes to the neutron-rich side of nuclear chart. Moreover, the shell gaps at $N=50$ and $N=82$ are supposed to gradually decrease with the decreasing of the proton-to-neutron ratio. These gaps are still sizable for these extreme neutron-rich light nuclei when the SO potential is enhanced due to the high isospin asymmetry, i.e., $\kappa_{so}=-\kappa$. Recently reported magic numbers, like $N = 14$, 16, 32, and 34, have been understood in terms of the monopole nucleon-nucleon interaction within the interacting shell model. Interestingly, one can also find the positive signals on the emergence of these new magic numbers from the present single particle model, which implies that the aforementioned monopole interaction can be absorbed in the good mean-filed potential. The single-particle spectra, under such a simple scheme, may not only provide a reasonable starting point for the community of shell model calculations in the continuum. but also serve for exploring the pairing correlation and nuclear deformation.

\begin{acknowledgments}
The authors appreciate Chong Qi for fruitful discussions and suggestions. This work is supported by the National Natural Science Foundation of China (Grant Nos. 12075121 and 11605089), and by the Natural Science Foundation of Jiangsu Province (Grant No. BK20150762 and BK20190067).
\end{acknowledgments}

\bibliography{myref}

\end{document}